\documentclass[12pt,preprint]{aastex}

\newcommand\de{\delta}

\newcommand\ta{\tau}

\newcommand\ph{\Phi}

\newcommand\bna{\mbox{\boldmath{$\nabla$}}}

\newcommand\<{\langle}
\renewcommand\>{\rangle}

\newcommand\fr{\frac}
\newcommand\tha{\theta}

\newcommand\half{\textstyle\frac{1}{2}}
\newcommand\ap{\approx}
\newcommand\cd{\cdot}

\newcommand\bk{\mbox{\boldmath{$k$}}}
\newcommand\br{\mbox{\boldmath{$r$}}}

\newcommand\bzero{\mbox{\boldmath{$0$}}}
\newcommand\by{\mbox{\boldmath{$y$}}}

\newcommand\bth{\mbox{\boldmath{$\theta$}}}
\newcommand\bde{\mbox{\boldmath{$\delta$}}}

\begin{document}
\title{Time delay by primordial density fluctuations: its biasing
effect on the observed mean curvature of the Universe}

\author{Richard Lieu$\,^{1}$ and Jonathan P.D. Mittaz$\,^{2}$}

\affil{\(^{\scriptstyle 1} \){Department of Physics, University of Alabama,
Huntsville, AL 35899.}\\
\(^{\scriptstyle 2} \){Cooperative Institute for 
Climate Studies, ESSIC, University of 
Maryland 2NOAA/NESDIS, Camp Springs, Maryland.}\\} 

\begin{abstract}

In this paper we specifically studied one
aspect of
foreground primordial matter density
perturbations: the relative gravitational time delay between a
pair of light paths converging towards an observer and
originating from two points on
the last scattering surface separated by the physical scale of
an acoustic oscillation.
It is found
that time delay biases the size of acoustic oscillations 
{\it systematically} towards
smaller angles, or larger harmonic numbers $\ell$, 
i.e. the mean geometry as revealed by CMB light becomes
that of an open Universe if $\Omega=1$.  
Since the effect is second order, 
its standard deviation $\delta\ell/\ell \sim (\delta\Phi)^2$
where $(\delta\Phi)^2 \sim 10^{-9}$ 
is the normalization of the primordial matter
spectrum $P(k)$, the consequence is
too numerically feeble to warrant
a re-interpretion of WMAP data.  If, however,
this normalization were 
increased to $\delta\Phi \gtrsim 0.01$
which is still well within the perturbation limit, the shift in the 
positions of
the acoustic peaks would have been substantial enough to implicate
inflationary $\Lambda$CDM cosmology.  Thus
$\Omega$ is not the only parameter (and by deduction inflation cannot be
the only mechanism)
of relevance to the understanding of {\it observed}
large scale geometry.  The physics that explains why $\delta\Phi$ is
so small also plays a crucial role,
but since this is a separate issue
independent of inflation, might it be less artificial to look for
an alternative solution to the flatness problem altogether?

\end{abstract}


\section{Introduction}

Among the so-called `secondary' physical mechanisms that re-process the 
CMB anisotropy, such as gravitational 
lensing, time delay, and the Sunyaev-Zel'dovich effect, time delay by
foreground inhomogeneities in the matter distribution
appears to be investigated least.  In the recent period a detailed 
treatment of the problem was provided by Hu \& Cooray (2001,
hereafter HC01), although
the general framework for calculating gravitational perturbation effects
on the CMB as published in the
review of CMB lensing by Lewis \& Challinor (2006) could
also be employed to carry the study further.

In HC01 the authors found an infra-red
logarithmic divergence
in the variance of the {\it absolute}
(or total) time delay along a randomly chosen direction
to the LSS.  
They `renormalized' this infinity by subtracting the
contribution from the `monopole' term of the matter power spectrum,
corresponding to the removal of a constant
uniformly across the sky.  Nevertheless, the remaining (finite) quantity
still carries the divergence trend.  More precisely the variance is dominated
by long wavelength fluctuations, prompting HC01
to consider it as an effect of very large coherence length, $\Delta \ell
\approx$ 2 which does not affect our interpretation
of the CMB anisotropy, because the net outcome 
is simply a gentle and arbitrary
distortion of the spherical shape of the LSS, completely negligible
over the size of one (or a few) cycles of CMB acoustic oscillations.

The coherence length inferred by HC01 should
be viewed with some caution, however, because the large delay
excursion of $\sim$ 1 Mpc/c
calculated there, which involved only the {\it zeroth} order
term, the path integral of the perturbing potential itself,
stems from the part of the matter power spectrum which 
carries the scale-invariant Harrison-Zel'dovich dependence
$P(k) \sim k$, i.e. there is a danger that the large
coherence length may simply be due to the infra-red divergence of the variance
rather than any genuine physical scale in the matter
spectrum.  To find the coherence length of
relevance to the question of CMB anisotropy distortion, it is necessary to
pursue the perturbation expansion to the next two
orders.  Not only are both results free from divergences, but also
it is only through a comparison of these two terms 
that the true coherence scale for variations in the {\it relative}
delay between two light paths separated by a small angle $\theta$
would become transparent.  We shall find that this scale
is defined by a physically significant parameter,
viz. the characteristic wavenumber
at which departures of $P(k)$ from the Harrison-Zel'dovich behavior
occurs for the first time.  The consequence is that appreciable
distortion of the LSS radius, with both amplitudes 
and wavelengths on par with the dimension of the primary acoustic oscillations
at the time of last scattering, can exist in principle.

\section{Perturbation in the gravitational potential from the 2dFGRS/WMAP1 power spectrum of primordial matter}

Although in the $k \rightarrow$ 0 limit the matter 
power spectrum has the form $P(k) \sim k$, the behavior
of $P(k)$ at large $k$ is more
complicated than an exponential cutoff.  It is possible, however, to
break down {\it any} general $P(k)$ into constituent terms, each of
the form $a_i ke^{-b_i k}$, and sum up 
the time delay fluctuation contributions from all the terms,
because $P(k)$ has the meaning of a variance, i.e. it too is
additive.  We may therefore write
\begin{equation} P(k) = A (a_1 k e^{-b_1 k} + a_2 k e^{-b_2 k} + \cdots),~{\rm with}~
\sum_i a_i = 1. \end{equation}
This empirical representation of $P(k)$ is not the same as the
more commonly used ones (e.g. Efstathiou, Bond, and
White 1992) but, as shall be seen in section 3,
the exponential form reveals coherent length scales of
foreground effects in a transparent way; in any case, provided our
formula for $P(k)$ fits the observational data (see below) the detailed
structure of the terms used to model the spectrum
is of no significance.

The resulting
value of $A$ in Eq. (1) that we shall obtain is
\begin{equation} A = 3.276 \times 10^7~~{\rm Mpc}^4, \end{equation}
and leads, by Eq. (A-6), to 
\begin{equation} \delta\Phi \approx 3 \times 10^{-5}, \end{equation}  
for a $\Omega_m =$ 0.3, $\Omega_{\Lambda} =$ 0.7, and $h =$ 0.7 cosmology
(Bennett et al 2003, Spergel et al 2007).
This
agrees well with the CMB temperature modulation of 
$3 \delta T /T$
at small $k$ as
measured by WMAP (see e.g. Bennett et al 2003), as it ought to,
because from Eq. (A-5)
\begin{equation}
(\delta\Phi)^2 = \lim_{k\to 0} \frac{d\Phi_k^2}{d{\rm ln}k} = 
\lim_{k\to 0} \left(\frac{3\delta T_k}{T_k}\right)^2,
\end{equation}
where the final step is explained in the material around Eq. (18.14)
of Peacock (1999).  The consistency between $\delta\Phi$ as derived from
our $z=$ 0 matter spectrum and the large scale CMB anisotropy re-assures us
that any corrections  we ignored, such as
the effect of vacuum domination at $z \lesssim$ 0.3, are
indeed minor.

If $P(k)$ has the simple form involving only the first term of Eq. (1)
with $a_1 =$ 1,
we may work out from Eq. (A-5) the correlation function for the perturbing
Newtonian potential $\Phi$ that arises from the linear growth of
primordial density contrasts, as
\begin{equation} \langle\Phi({\bf r})\Phi({\bf r'})\rangle = \frac{9\Omega_m^2 H_0^4}
{32\pi^3} \int \frac{d^3 {\bf k}}{k^4} 
e^{i\bk\cd(\br-\br')} P(k) = \frac{(\delta\Phi)^2}{4\pi} \int 
\frac{d^3 {\bf k}}{k^3} e^{i\bk\cd(\br-\br')} e^{-bk}, \end{equation}
where $\Phi$ is assumed to be time-independent and Gaussian distributed,
and in the final step use was made of Eqs. (A-4) and (A-6).  If $P(k)$ is
given by the full Eq. (1) instead, 
then the rightmost side of Eq. (5) will be a sum
of similar terms, each carrying the
exponent $b_i$ and with $(\delta\Phi)^2$ replaced by
$a_i (\delta\Phi)^2$.

Implementing now the observed power spectrum, the 
most up to date data are from the 2dFGRS galaxy survey
(Cole et al 2005)
after they are deconvolved and aligned with the WMAP1 normalization
by setting the $\sigma_8^2$ parameter  to $\sigma_8^2 =$ 0.74
(Sanchez et al 2006).  We found that the resulting dataset can adequately
be fitted with a function for $P(k)$ of the form given by Eq. (1)
and involving three exponential terms, with
the value of $A$ as already quoted in Eq. (2) and the values of $a_i$
and $b_i$ ($i =$ 1,2,3) as shown in Table 1.  This best-fit
spectrum, which closely follows that of WMAP1's
$\Lambda$CDM model (Spergel et al 2003) is plotted in Fig. 1.

\clearpage
\begin{figure}[H]
\begin{center}
\includegraphics[angle=270,width=4in]{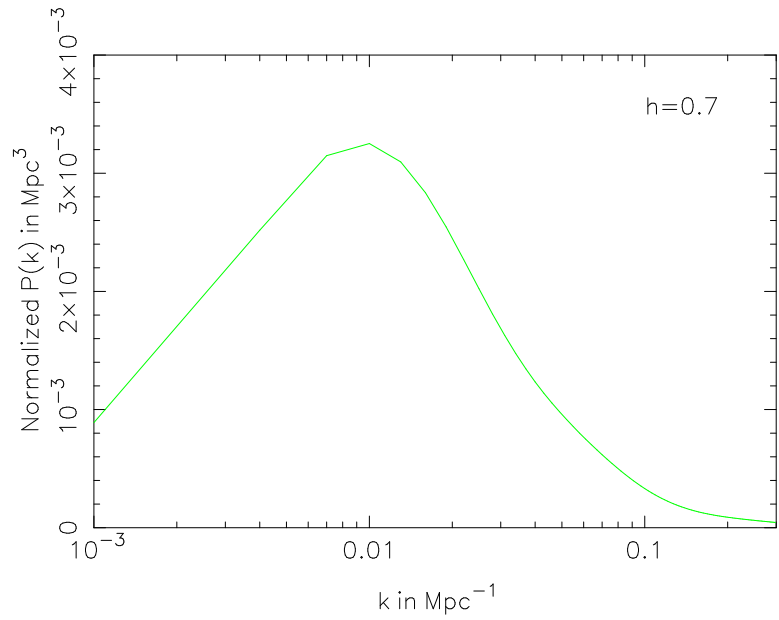}
\vspace{-4mm}
\end{center}
\caption{The best $P(k)$ model of the WMAP1
normalized 2dFGRS data as given by Eq. (1) with
$A=1$ and the remaining parameters as shown in
Table 1. The Hubble constant assumed is $h =$ 0.7.}
\end{figure}

\clearpage



\section{Cosmological time delay variation from primordial density contrasts}

We proceed towards calculating the excursion in time delay along two
light paths of equal lengths.
Our starting point is a flat Universe having its average density at
the critical value, as required by the WMAP observations (Bennett et al
2003, Spergel et al 2007).
We employ Cartesian comoving coordinates and the 
conformal time coordinate $\eta$, so that in a
perturbed flat FRW space light propagates along the null
geodesics of the metric
 \begin{equation}
 ds^2 = (1+2\ph)d\eta^2 - (1-2\ph)(dx^2+dy^2+dz^2),
 \end{equation}
as though the expansion factor $a(\eta)$ plays no role.
Let light signals arrive at the observer's origin from a source at
$(x,0,0)$.  Although the signal is coming towards us, we can,
by optical reciprocity,
solve the equation starting at $x=0$ and following its path 
backwards.

\clearpage
\begin{table}
\caption{Parameters (with 1-$\sigma$ errors)
for the best $P(k)$ model of the WMAP1
normalized 2dFGRS data ($h =$ 0.7).
The form of the model is given by Eq. (1).
Three exponentials were needed to fit the data.}
\begin{tabular}{lccr}
$a_i$ &  $a_i$ error & $b_i$~(Mpc) & $b_i$ error~(Mpc) \\
\hline
0.883 & 0.17  & 128.0 &  11.0 \\
0.114 & 0.02  & 39.3  &  3.80 \\
0.0047 & 0.001 & 9.04 &  0.80 \\
\hline
\end{tabular}
\end{table}
\clearpage

Suppose a signal arrives from a direction making a small 
angle $\bth$ w.r.t. the $x$ axis.  This may correspond to some 
`off-axis' point
on the LSS in the case of the CMB.  The
conformal time of travel for our light signal is, from Eq. (6),
 \begin{equation} \eta=\int_0^x[1-2\ph(x',\bth x')]dx', \end{equation}
with an ensuing time delay of
\begin{equation} \ta=\int_0^x\,2\ph(x',\bth x')dx', \end{equation}
where $x$ and $x'$ can be present day physical distances 
if we set $a(z=0) = a_0=1$ (here we ignored the geometric time delay,
with is comparatively negligible as pointed out by HC01).   Moreover, if
we use the current value of $\Phi$, as in the previous section, to calculate
$\tau$, then $\delta\tau = \delta t/a(t) = \delta t$.
The `absolute' variance of the time delay
along a random direction $\bth$ to the LSS 
\begin{equation} \<\ta^2(\bth)\> = \int_0^x 2dx'\int_0^x 2dx'' \<\Phi(x', \bth x')
\Phi (x'', \bth x'') \>, \end{equation}
is logarithmically divergent because the quantity $\<\Phi({\bf r})
\Phi ({\bf r'}) \>$ is, by Eq. (5), of the form $\int_0^k dk'/k'$ 
at small $k$.  As stated in the beginning of the paper, HC01 renormalized
$\<\ta^2(\bth)\>$ by subtracting an infinite constant from it.  

Of more relevance to understanding the CMB acoustic peaks is
the relative delay in the arrival time between the above signal and another
light signal emitted simultaneously from the same distance, but along the
`on-axis' direction $\bth= 0$, i.e.
 \begin{equation} \ta(\bth)-\ta(\bzero)=
 \int_0^x\,2x'\bth\cd\bna\ph(x',\bzero)dx' + \int_0^x\,{x'}^2 (\bth\cd\bna)^2
 \ph(x',\bzero)dx' + \cdots, \end{equation}
where $\bna$ is the gradient operator transverse
to the vector ${\bf x}$.  
assuming for the time being that the form of $P(k)$ is given by Eq. (A-4),
we can construct the correlation function with the help of Eq. (5), as
\begin{eqnarray}
 \<\nabla'_i\ph(\br')\nabla''_j\ph(\br'')\> &=&
 \fr{(\de\ph)^2}{4\pi}
 \int \fr{d^3\bk}{k^3}k_ik_je^{i\bk\cd\br}e^{-bk}\nonumber\\
 &=&\fr{(\de\ph)^2}{r^2}\left\{\de_{ij}\left[1-\fr{b}{r}
 \arctan\left(\fr{r}{b}\right)\right]
 +\fr{r_ir_j}{r^2}\left[\fr{3b}{r}\arctan\left(\fr{r}{b}\right)-2
 -\fr{b^2}{r^2+b^2}\right]\right\},\nonumber\\
\end{eqnarray}
where 
 \begin{equation} \br=\br'-\br'', \end{equation}
and the indices $i,j$ denote {\it two} orthogonal components in directions
transverse to $x$.
This enables us to derive the lowest order
term for the variance in the relative time delay
$[\delta\ta (\theta)]^2 = \<[\ta(\bth)-\ta(\bzero)]^2\>$, viz. the
first term on the right side of Eq. (10), as
 $$ [\delta\ta (\theta)]^2 =
 (\de\ph)^2\bth^2\int_0^x\,2x'dx'\int_0^x\,2x''dx''
 \left[\fr{1}{r^2}-\fr{b}{r^3}\arctan\left(\fr{r}{b}\right)\right], $$
where $r=|x'-x''|$, consistent with Eq. (12).

Transforming now to
the new variables ${\bar x}=(x'+x'')/2$ and
${\tilde x}=x'-x''$, the resulting
integrand is symmetric in $\tilde x$, so we can
restrict the range of $\tilde x$ to positive values,
introducing an extra factor of 2.  Thus
 \begin{eqnarray}
[\delta\ta (\theta)]^2 &=&  2(\de\ph)^2\bth^2
 \int_0^x d\tilde x\int_{\frac{\tilde x}{2}}^{x-\frac{\tilde x}{2}}d\bar x
 \fr{4\bar x^2-\tilde x^2}{\tilde x^2}
 \left[1-\fr{b}{\tilde x}\arctan\left(\fr{\tilde x}{b}\right)\right]\nonumber\\
& = & \fr{4}{3}(\de\ph)^2\bth^2\int_0^x d\tilde x
 \fr{2x^3-3x^2\tilde x+\tilde x^3}{\tilde x^2}
 \left[1-\fr{b}{\tilde x}\arctan\left(\fr{\tilde x}{b}\right)\right].
 \end{eqnarray}
By restricting ourselves to the limit $x \gg b$ 
((appropriate to emission distances
$x \gg b_1 \approx$ 0.1 Mpc - see Table 1 - the CMB LSS clearly satisfies
this criterion),
it becomes
straightforward to complete the calculation, because only
one term stands out.  The result is  
\begin{equation} 
[\delta\ta (\theta)]^2=\fr{4}{3}(\de\ph)^2\bth^2
\frac{x^3}{b}\arctan\left(\fr{x}{b}\right) = 
\fr{2\pi}{3}(\de\ph)^2\bth^2 \fr{x^3}{b}.
\end{equation}
Note that according to Eq. (14) the variation in the time
delay difference between two points A and B on the LSS subtending an 
angle $\bth$ at the observer O is $\delta\ta (\theta) \sim \theta$.
This behavior indicates that we are in the regime of {\it coherent
delay}, i.e. provided $\theta$ is sufficiently small the two rays
sampled a primordial matter potential gradient which may  be
regarded as constant.  Thus, if in Figure 2a a third point S$_3$ on the LSS is 
collinear with two other points S$_1$ and S$_2$,
then a constant gradient would
imply equality between the `S$_1$ to S$_2$'
and `S$_2$ to S$_3$' relative delays, i.e. both are
$\delta\ta (\theta)$.  Consequently the net S$_1$ to S$_3$ delay will be
$2\delta\ta (\theta)$.  Since 
the angular separation between S$_1$ and S$_3$
is $2\bth$ at O, we have $\delta \ta(2\theta) =
2\delta\ta (\theta)$, fully
consistent with the 
$\delta\ta (\theta) \sim \theta$ dependence.
\clearpage
\begin{figure}[H]
\hspace{-1cm}
\includegraphics[angle=0,width=4in]{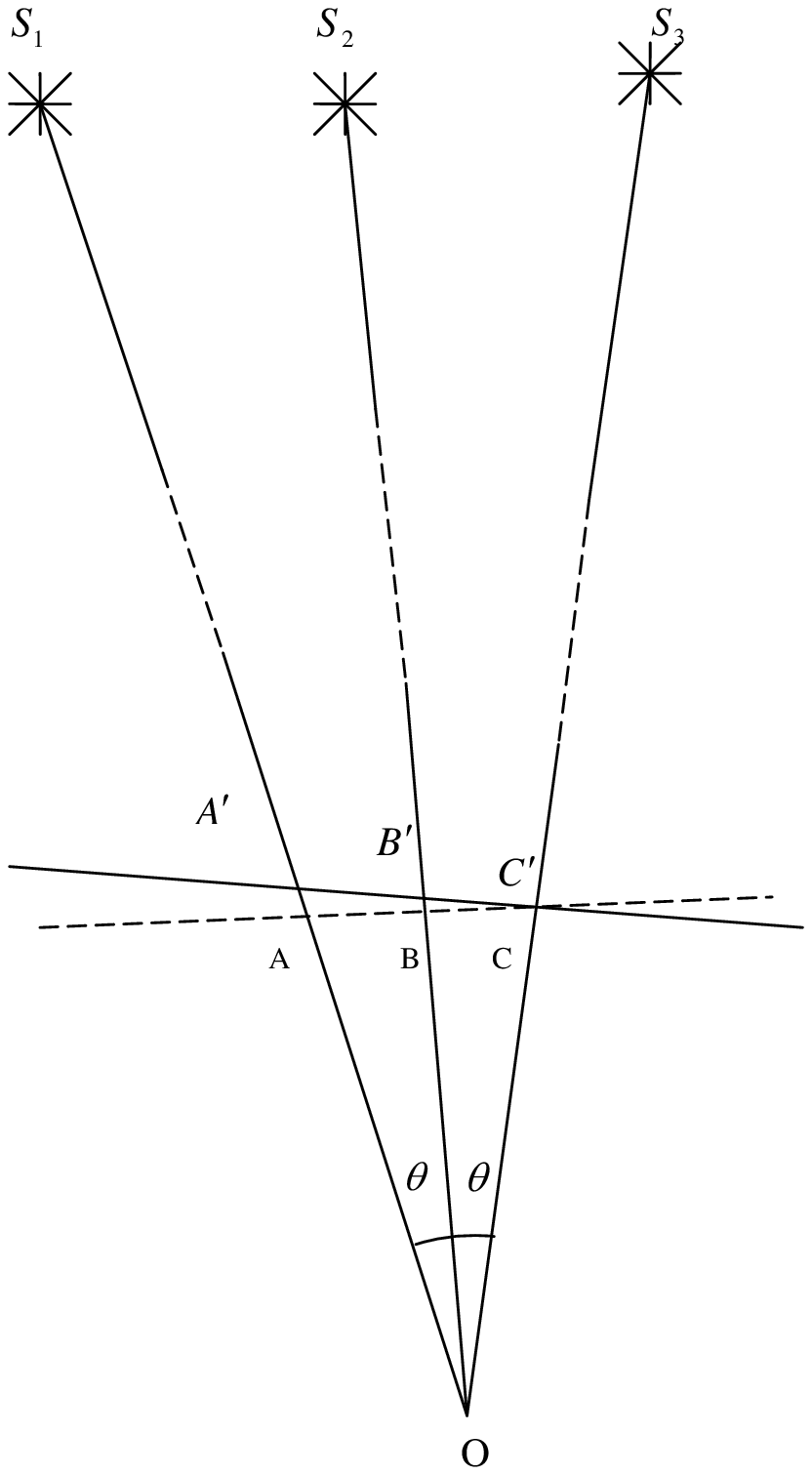}
\hspace{-2.5cm}
\vspace{3cm}
\includegraphics[angle=0,width=4in]{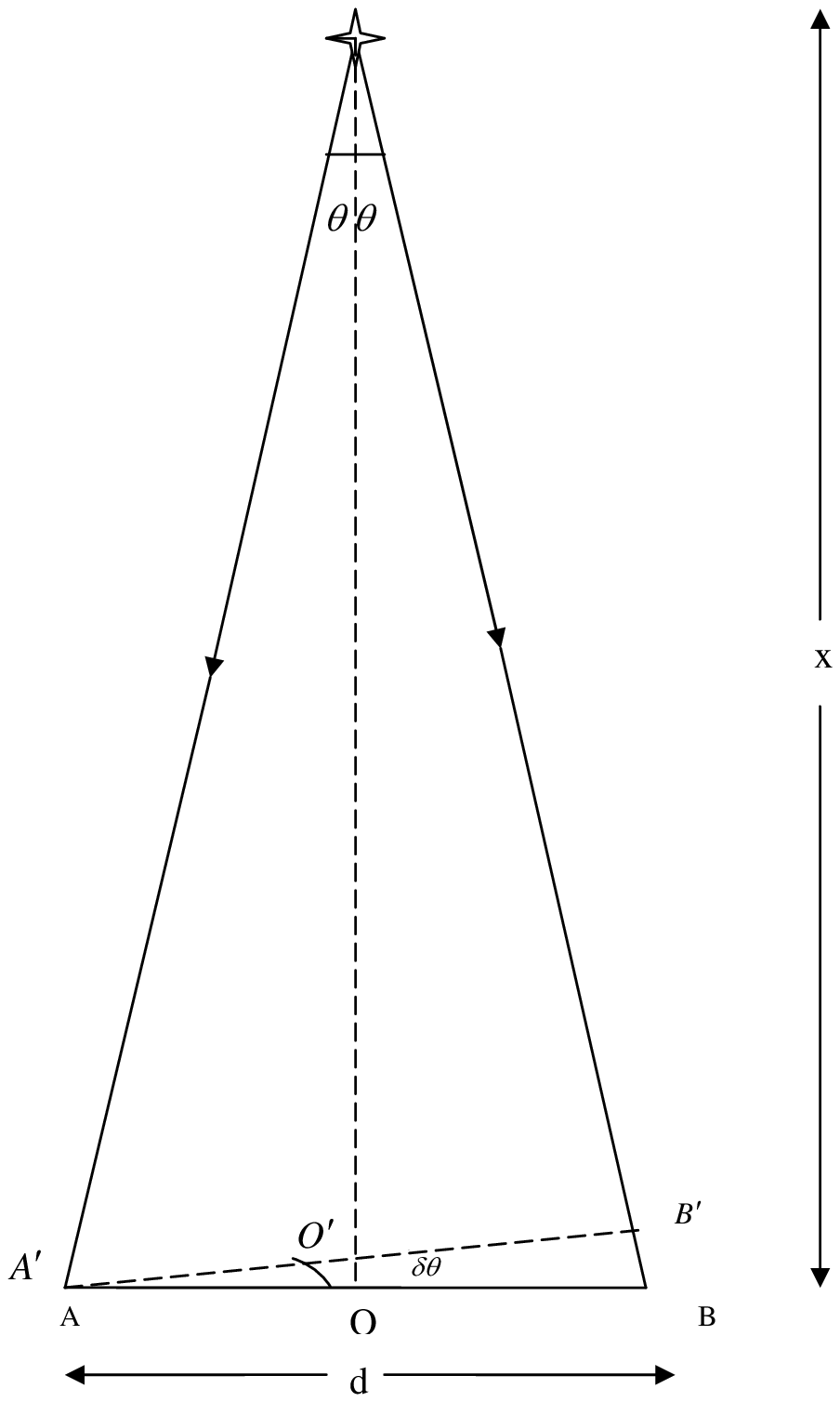}
\vspace{-8cm}
\caption{Illustrating the phenomenon of coherent time delay.
{\it 2a}: simultaneously emitted light signals from three 
symmetrically located and equally 
distant sources would have arrived at points A, B, and C also
at the same instance if there were no gravitational time delay.
Provided $\theta$ is small enough that only the first spatial
derivative of the potential is responsible for any delay, then
the signals from S$_1$ and S$_2$ will e.g. have reached
A' and B' when that from S$_3$ is at C'$=$C, where
AA' $=$ 2BB'.  The result is a net displacement of the angular position
of all three sources by the same amount to the right.
{\it 2b}: If we reverse the light paths by letting the observer be
a single source and the former (S$_1$, S$_2$, S$_3$) be replaced by
(B, O, A), then the source will appear shifted
in the opposite direction (i.e.
to the left) w.r.t. the observer whose telescope aperture is AB,
as it should do.}
\end{figure}
\clearpage

Turning to the $\nabla^2\Phi$ term, 
which takes into account the difference in the
potential gradient between the two rays, its
exact mathematical form
can be derived directly from the last integral of Eq. (10), but
in this paper we will present instead
another method of calculation which also provides unusual
insights to the relationship between time delay and 
gravitational lensing.
Such material is to be found in sections 4 and 5.
For now we simply give the result, i.e. an
expansion of $[\delta\ta (\theta)]^2$ to 
include the $\nabla^2\Phi$ term.  It is
\begin{equation}
[\delta\ta (\bth)]^2 \approx
\fr{2\pi}{3}(\de\ph)^2\theta^2 \fr{x^3}{b} \left(1 - \frac{1}{20}
\frac{x^2 \theta^2}{b^2} \right)~{\rm for}~\theta < \theta_m;
\end{equation}
or, if $P(k)$ has the more general form of Eq. (1), the contribution from
the $i$th term to the variance (note the label $i$ here has a 
{\it different} meaning from that in Eq. (11)) will be
\begin{equation}
[\delta\ta_i (\bth)]^2 \approx
\fr{2\pi}{3}(\de\ph)^2\theta^2 x^3
\frac{a_i}{b_i} \left(1 - \frac{1}{20}\frac{x^2 \theta^2}{b_i^2}\right)
~{\rm for}~\theta < \theta_m^i,
\end{equation}
where in Eqs. (15) and (16) the parameters $\theta_m$ and $\theta_m^i$
are yet to be defined below - see Eqs. (17) and (20).
From Eq. (15) we see that the higher order ($\theta^4$) term, once
it assumes importance, will halt the linear rise of $\delta\ta (\theta)$
with $\theta$.  
Eventually, 
when $\theta$ becomes large enough, all
the higher order terms of Eq. (10) will take their place
to ensure that $[\delta\ta (\theta)]^2$ reaches constancy\footnote{Strictly
speaking there may be a gentle rise with $\theta$ because of the logarithmic
divergence problem mentioned in section 1.  This has no significant
impact on any
of the results presented in this paper, however.},
as it must do, because two widely separated rays are {\it uncorrelated}.
Thus, once the $\theta^4$ term is no longer negligible, {\it incoherence}
takes over.
A reasonable (and conservative)  way of estimating the maximum
(or `plateau') value of $[\delta\ta (\theta)]^2$ is to 
find the angle at which
the first $\theta$-derivative of the right side of Eq. (15) vanishes.  This
occurs when
\begin{equation} \theta_m = \sqrt{10}\frac{b}{x},  \end{equation}
at which point $\delta\tau$ reaches the constant (saturation) value of
\begin{equation} 
[\delta\ta (\bth)]^2 =
[\delta\tau (\theta_m)]^2 = 
\frac{10\pi}{3} (\delta\Phi)^2 xb~{\rm for}~\theta \geq \theta_m 
\end{equation}
Here $\theta_m$ may be defined as the {\it coherence angle} for
time delay.  It is also closely related to the expansion
parameter of the perturbation expansion of Eq. (15), because
the ratio of the second order to the first order term of Eq. (15)
is $2\theta^2_m$.   When a more general form of $P(k)$, like Eq. (1), 
is considered, we have
\begin{equation}
[\delta\ta_i (\bth)]^2 = [\delta\tau_i (\theta_m^i)]^2 =
\frac{10\pi}{3} (\delta\Phi)^2 x a_i b_i~{\rm for}~\theta \geq \theta_m^i,
\end{equation}
where
\begin{equation}
\theta_m^i = \frac{\sqrt{10} b_i}{x}
\end{equation}
for each component $i$ of $P(k)$ in Eq. (1).  The total variance of
time delay is then the sum over $i$ of all the individual variances
$[\delta\ta_i (\bth)]^2$ as given by Eqs. (16) and (19).
Obviously, for rays separated by angles $\theta > \theta_m$  the
time delay is always completely incoherent (or random).
As will be demonstrated in the next section, the significance
of the transition angle $\theta_m$ is that
for separation $\theta < \theta_m$, the
variance for the coherent (or systematic) and
incoherent delay
is given by the first and second
order terms of Eq. (15) respectively.

\clearpage
\begin{figure}[H]
\begin{center}
\includegraphics[angle=90,width=4in]{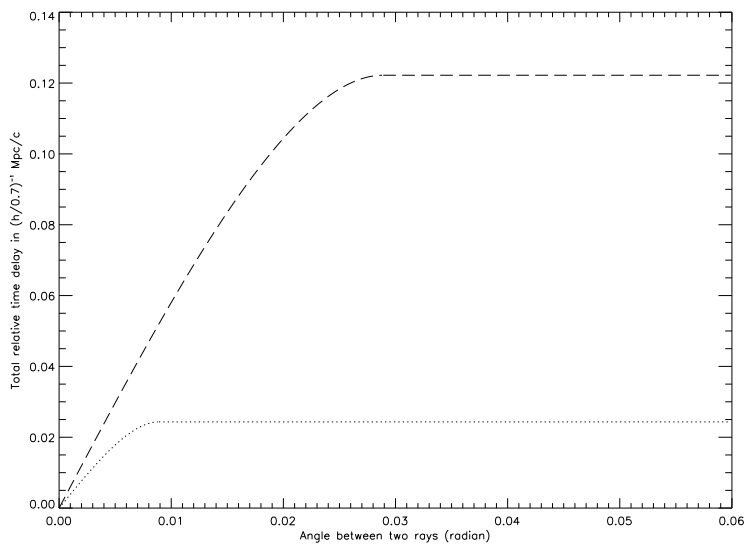}
\vspace{0.0mm}
\end{center}
\caption{1-$\sigma$
relative time delay $\delta\ta^i(\theta)$ between two
rays from the LSS due to the various terms $i$ of primordial
density perturbation of Eq. (1).
In the limit 
$\theta \leq \theta_m^i$ where $\theta_m^i$ given by Eq. (20),
$\delta\ta^i(\theta)$ is given by Eq. (16).  When
$\theta > \theta_m^i$, $\delta\ta^i(\theta)$ saturates to the
value in Eq. (19).
Long dashes correspond to the first term $i=1$, and the dotted line to
the second term $i=2$ (the $i=3$ term was ignored due to its insignificance).
Values of
$a_i$ and $b_i$ are shown in Table 1.}
\end{figure}
\clearpage

How does time delay affect CMB observations?  When
light signals emitted
during the same redshift $z = z_{{\rm LSS}}$
from two points on the LSS which subtend the angle $\theta$ at the
observer are to arrive simultaneously {\it and yet} the times of flight are
delayed w.r.t. each other, the signals would necessarily have covered
different distances - the slower one
must have undertaken a shorter journey.  As illustrated in Figure 4,
the consequence is a `tilt' of the LSS in that vicinity.  Hence, an 
angle $\theta$ which normally corresponds to an anisotropy
on the LSS spanning the comoving distance $x\theta$ would now be randomly
mapped to a {\it larger} distance $x(\theta + \delta\theta)$
where, from Figure 4,
\begin{equation} 
\frac{\delta\theta}{\theta} = \frac{\delta\ell}{\ell} =
\frac{[\delta\tau(\theta)]^2}{2x^2\theta^2} =
\frac{1}{2x^2\theta^2} \sum_i [\delta\tau_i (\theta)]^2,
\end{equation}
with 
the contribution to $(\delta\tau_i)^2$  from the $i$th term of
Eq. (1) given by Eqs. (16) and (19).  The net outcome is a {\it systematic}
shift of the CMB acoustic peaks, which are features with fixed
comoving scales, towards smaller angles or higher harmonic numbers $\ell$.

It should be mentioned that another possible test of cosmological time delay
concerns two light paths propagating through different parts of
the Universe, but connecting the same source with the observer, i.e.
a strong gravitational lensing scenario, under which multiple images occur.
If this source undergoes flaring behavior, a time lag between the
light curves of a pair of multiple images might result from
perturbations in the distribution of primordial matter.  In practice, however,
this is a sensitive test only if the lowest order ($\delta\ta (\theta) \sim
\theta$) term contributes to the delay.  As it turns out, this term
does {\it not} play a role
because the potential time delay we have hitherto
been considering is cancelled by another effect - the geometric delay -
which {\it is} important under this scenario (see Bar-Kana 1996
and Seljak 1994).
Since the
higher order terms are too feeble to be measured, no
meaningful constraints can be provided here.

\clearpage
\begin{figure}[!h]
\hspace{4cm}
\includegraphics[angle=0,width=3in]{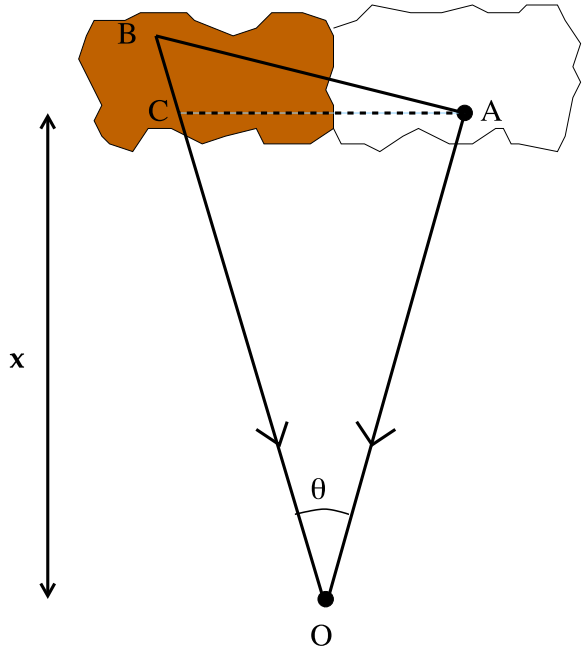}
\caption{The effect of CMB time delay on the 
appearance of the acoustic peak hot and cold
spots.  Two proximity locations on the observed LSS are the points A and B.
Were time delay absent, they would have been the points A and C.
Here the distance BC is typically $\delta\tau(\theta)$, the quadrature
sum (over all $i$) of the $\delta\tau_i(\theta)$ contributions
as given by Eqs. (16) and (19), because it
is the relative delay between the light signals
emitted at A and C, with the path AO suffering from more delay than BO
in this case.  Since any LSS plane samples the same distribution of
hot and cold spots as any other, the tilting of the LSS by time delay
invariably means an observer charted slightly larger physical distance on
the LSS for a certain angular separation between two temperature sensors.
This {\it systematically} causes all acoustic peaks to move towards
smaller $\theta$, or larger $\ell$, in the manner of Eq. (21).}
\end{figure}
\clearpage

\section{Deflection angle fluctuation from primordial density contrasts}

A significant gain in our general
understanding of foreground effects
is afforded by examining also the question of CMB lensing.
In addition to the $x$-axis of section 3, which points along a radial
direction from the observer's origin back to the LSS, we now introduce
two more Cartesian (comoving) coordinates $\by$, both measuring distances
transverse to the light path.  Tracing the path towards the LSS as before,
the equation of motion for a light signal is
 \begin{equation} \fr{d^2\by}{dx^2}=-2\bna\ph. \end{equation}
Let a signal arrive from some direction which makes a (small) angle $\bth$
with the $x$ axis.  Then the solution to (28) is
\begin{equation} 
 \by=\bth x+\bde\by=
 \bth x-2\int_0^x dx'\,(x-x')\bna\ph(x',\bth x'),
\end{equation}
with appropriate initial conditions.

From Eq. (30) emerges
that the correlation between the values of the
deviation angle $\bde\by/x$ for different values of $\bth$,
 \begin{eqnarray}
 C_{ij}(|\bth'-\bth''|)&\equiv&\fr{1}{x^2}
 \<\de y_i(\bth')\de y_j(\bth'')\>\nonumber\\
 &=&\fr{4}{x^2}\int_0^xdx'(x-x')\int_0^xdx''(x-x'')
 \<\nabla'_i\ph(x',\bth'x')\nabla''_j\ph(x'',\bth''x'')\>.
 \end{eqnarray}
By means of Eqs. (24) and (11), one can expand the ensuing
correlation function $C(\bth)$ as a Taylor series (see Appendix B
for details on intermediate steps), i.e.
\begin{equation}
C(\bth) = C_{ii}(\bth) \equiv 
\frac{\<\de y_i(\half\bth)\de y_i(-\half\bth)\>}{x^2}
= \frac{4\pi}{3} (\delta\Phi)^2 \frac{x}{b} \left(1-\frac{1}{20}
\frac{x^2\tha^2}{b^2} + \cdots \right),
 \end{equation}
where the sum represented by a repeated index is over the two
transverse dimensions.
 


\section{Reciprocity of light propagation: the relationship between time delay and deflection}

An interesting connection exists between sections 3 and 4, in
that the key formulae derived in each section, viz.
Eqs. (15) and (25), are closely related to
each other because of a physical reason - the
reciprocity of light propagation.

We first consider the scenario of a source S emitting light signals
that enter an observer's circular
telescope aperture at the extremeties A and B
such that the angle ASB is $2\tha$, see Fig  2b.
Let us assume that when the Universe is homogeneous, the wavefronts
are parallel to the aperture as they arrive, i.e. the signals reach
points A and B simultaneously.
We further suppose that the same is true for two points C and D on the
extremeties of another aperture diameter 
perpendicular to the AB line - the source is `on-axis'.

In the presence of primordial matter density perturbation, let
the signal propagating in the SB direction reach the point B' when
the SA signal has already arrived at A, i.e. the SA signal
suffered from a smaller delay.  The outcome is a tilting of the
wavefronts as they enter the aperture - the source is now seen to
have moved `off-axis' along the AB line to a new position leftwards
of the center O.  The angle of tilt is $\delta\tha = 2\delta\ta (\tha)/d$
in the limit of coherent time delay (i.e. small $\theta$)
where $\delta\ta(\tha)$ is the distance BB' and $d$ is the diameter AB.
Repeating our above argument to the points C and D, and noting that
the relative delay between the SC and SD directions is {\it independent}
of that between SA and SB, we realize that the variance
\begin{equation}
(\delta\bth)^2 = \frac{8[\delta\ta (\bth)]^2}{d^2}
\end{equation}
applies to the overall shift in the
position of the source on the two dimensional sky.  Of particular
interest is the fact that for a given source distance $x$ the
quantity 
\begin{equation} \tha = \frac{d}{2x} \sim d. \end{equation}
Thus, in order for the positional
shift to be the same amount irrespective of aperture size $d$ we
must have $\delta\ta (\bth) \sim \tha$ where $\tha^2 = \bth \cdot \bth$.
From section 3 we saw that this condition holds only when $\tha$, hence
$d$, is small.

To link sections 3 and 4, however, we have to consider a second
scenario, under which the distribution of primordial matter remains
the same as before, but
A, O, and B are now three simultaneously emitting
sources and S is the observer (equipped with a small telescope).
This reversal of the light paths converts Figure 2b into Figure 2a.
By the reciprocity of light propagation, the signals must arrive
at S in such a way that the change in
the positions of the three sources as perceived by S after
the `turning on' of the matter perturbation involves an angular shift
by the same amount $-\delta\bth$ for each source.
Yet according to section 4, this shift has a variance given precisely
by the correlation function $C(\bth)$, 
where $\bth$ is the angle the two sources subtend at
the observer S, viz.
\begin{equation}
(\delta\bth)^2 = \frac{\<\de y_i(\half\bth)\de y_i(-\half\bth)\>}{x^2}
=  C(\bth),
\end{equation}
with $C(\bth)$ being given by Eq. (25).

If we now take the limit
$\tha \rightarrow 0$.  By Eqs (28) and (22), the latter in
section 4, we have
\begin{equation}
(\delta\bth)^2 = \fr{\<\de\by^2\>}{x^2} = 
\frac{4\pi}{3} (\delta\Phi)^2 \frac{x}{b},
\end{equation}
which is a constant shift (of O w.r.t. A, and B w.r.t. O) independent 
of $d$ and $\tha$.  Such a behavior is also consistent with the
requirement stated at the end of the previous paragraph when we
considered our first scenario.  Thus, from Eqs. (26), (27), and (29)
we deduce that
\begin{equation}
[\delta\ta (\theta)]^2=
\fr{2\pi}{3}(\de\ph)^2\bth^2 \fr{x^3}{b},
\end{equation}
which is the low $\theta$ limit of the variance in the relative time delay
between two light paths separated by angle $\bth$, as derived in
Eq. (15) of section 3.  Thus, it is now clear that sections 3 and 4
can be unified by the principle of light reciprocity.

Of even more interest, however, is the regime of larger 
$\tha$, where incoherence between the two rays becomes important.
Here we already saw from Eq. (25) of section 4 that $C(\bth)$,
hence $(\delta\bth)^2$ by Eq. (28), is no longer a constant, but
decreases away from constancy as $\tha$ increases.  This decrease is
expected, because when $\tha$ becomes sufficiently large the two light
paths are independently perturbed by {\it different} primordial 
density fluctuations, i.e. the correlation function
$\<\de y_i(\half\bth)\de y_i(-\half\bth)\>$, hence
$(\delta\bth)^2$ by Eq. (28),  assumes relatively small
values.  The two sources shift their positions randomly w.r.t.
each other, leaving behind little net systematic drift of their
centroid across the sky (according to S).
In fact, when we
compare Eq. (25) in full with Eqs. (27) and (28), we obtain an expression
for $[\delta\ta (\theta)]^2$ in complete agreement with Eq. (15).
This consistency provided an important cross-check of
the robustness of our analysis, and explains why the $C_2$ term
in the variance represents incoherence effects - it is precisely this
term that decorrelates $C(\bth)$.  To labor upon this point even
further, we observe that the variance of the
{\it random relative deflection}
between the two rays:
$$
\fr{1}{x^2}\<[\de\by(\half\bth)-\de\by(-\half\bth)]^2\> 
 =2[C(0)-C(\bth)],
$$
has the $C_2$ coefficient in is leading term.  Thus, it is completely
clear that while the $C_0$ coefficient concerns absolute deflection and
coherent delay, the physics of $C_2$ is relative deflection and incoherent
delay.

A word of caution, however, before we leave this section.  
Although the phenomena of time delay and deflection are two sides of
the same coin, the `flipping of the coin' involves reversing light path
arrows, i.e. for a given set of arrows
we should {\it not}
conclude that delay presents no new physical
effects than those ensuing from lensing.  In particular, for
the CMB which is emitted
from a three dimensional distribution of sources, a varying time delay on
simultaneously observed signals from different directions allows {\it depth}
to play a role in the problem - as is already explained towards
the end of section 3.
This depth effect cannot be reproduced in any way by lensing.


\section{Time delay distortion of the CMB acoustic peaks}

We finally return to the original subject of this paper, viz. the
degree to which time delay by foreground 
primordial matter re-processes the CMB primary anisotropy.
In Eq. (1) and Table 1, we consider only the first two terms in
our breakdown of density fluctuations, viz. $i=1,2$ of Eq. (1).
Their coherence
lengths are, by Eq. (20), $\tha_m^1=\tha_1=$ 0.0289 or 
$\ell_1 =$ 108.8, and
$\tha_m^2 = \tha_2=$ 0.00886 or
$\ell_2 = \pi/\tha_2 =$ 354.4.
Using Eq. (21), then, one derives the
gaussian smoothing width of CMB anisotropy power, due to the $i=1$
perturbation only, as
\begin{equation}
\frac{\delta\theta}{\theta} =  4.5 \times 10^{-8} \times
(\delta\Phi/3 \times 10^{-5})^2 \times \left\{
\begin{array}{ll}
\frac{\theta_1}{\theta} , & \theta > \theta_1,\cr
2(1 - 600\theta^2) , & \theta < \theta_1.\cr
\end{array}\right.
\end{equation}
Likewise one also derives the width due to the $i=2$ term as
\begin{equation}
\frac{\delta\theta}{\theta} = 1.9 \times 10^{-8} \times
(\delta\Phi/3 \times 10^{-5})^2 \times \left\{
\begin{array}{ll}
\frac{\theta_2}{\theta} , & \theta > \theta_2,\cr
2(1 - 6363\theta^2) , & \theta < \theta_2.\cr
\end{array}\right.
\end{equation}
The total width $\delta\theta/\theta$ at a given $\theta$ is then the
quadrature sum of the contributions (at that value of $\theta$) from
both the $i=1$ and $i=2$ terms, as in Eq. (21).

A graph of $\delta\ell/\ell = \delta\theta/\theta$ versus $\ell = \pi/\theta$
for the $i=1$, $i=2$
components and their total is shown in Figure 5.  We explained
in the end of section 3 that the smoothing kernel for $\delta\theta/\theta$
is a one-sided gaussian function in the case of time delay perturbations, i.e.
structures on the LSS are invariably biased towards having smaller
perceived angular sizes irrespective of how time delay may tilt the
LSS.  Specifically the kernel $K(\theta,\ell)$ that enters the
integrand for the lensed CMB temperature correlation function, viz.
Eq. (A5) of Seljak (1996) is
\begin{equation}
K(\theta,\ell) = \left\{
\begin{array}{ll}
e^{-\ell^2 \sigma^2(\theta)/2} , & 
\ell > \frac{\pi}{\theta},\cr
0  , & \ell < \frac{\pi}{\theta},\cr
\end{array}\right.
\end{equation}
where $0.6\sigma(\theta) = \delta\theta$ with $\delta\theta$ given by
Eqs (31) and (32) added in quadrature, and the factor of 0.6 is due
to the fact that the standard deviation of a
one-sided gaussian  having the usual form for its exponent has
standard deviation 0.6$\sigma$.  

Thus, after applying the (normal)
simplifying procedures of Limber
approximation and retention of only the isotropic term $J_0 (\ell\theta)$
in Seljak's (A5), one arrives at the lensed correlation function due
to time delay
\begin{equation}
\tilde C_{\ell} = \int_0^{\pi} \theta d\theta \int_0^{\infty} \ell d\ell
K(\theta,\ell) C_{\ell} J_0 (\ell\theta),
\end{equation}
which is to be contrasted with Eq. (A7) of Seljak (1996).
Owing to the
smallness of the effect as dictated by the numerical coefficients
in Eqs. (31) and (32), however, we did not convolve the acoustic
peaks according to the recipe of Eq. (34), because
the distortion is completely unnoticeable\footnote{In order for
any smearing effects to distort the acoustic peaks to the point that the
resulting standard cosmological model no longer matches WMAP3 data,
it is necessary for $\sigma(\theta)/\theta$ to be at least 10 \% at the
center of the first peak.}.  We can therefore corroborate HC01,
after this comparatively more detailed and specific investigation
(see section 1),
by concluding that relative time delay by foreground primordial density
fluctuations does not lead to any appreciable changes in the prediction
of the standard cosmological model on the CMB acoustic anisotropy.

\clearpage
\begin{figure}[H]
\begin{center}
\includegraphics[angle=90,width=4in]{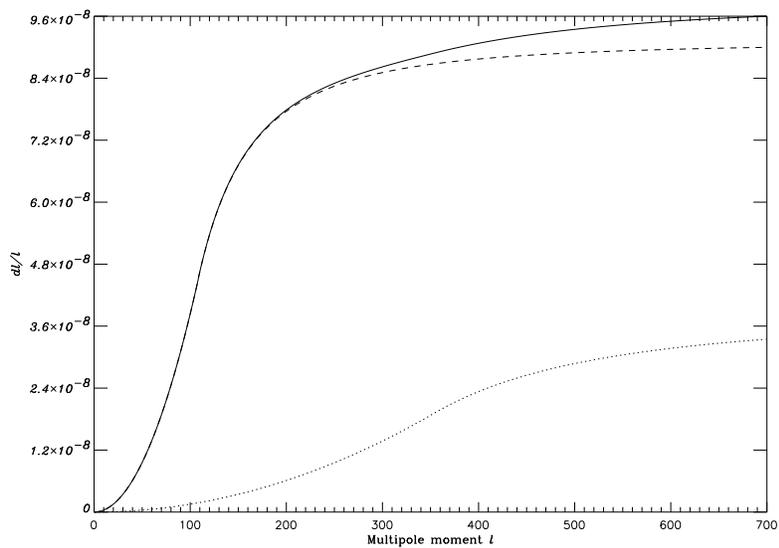}
\vspace{0cm}
\end{center}
\caption{The percentage variation of the size of an acoustic
harmonic $\ell$  due to
time delay differences along the LSS, i.e. Eqs. (31) and (32), is
plotted as a function of $\ell$ for
contribution from the mode $i=1$ (long dashes) and $i=2$ (short dashes)
of primordial density fluctuation in Eq. (1) and Table 1.  The
total $\delta\ell/\ell$ curve, obtained by adding the two dashed lines
in quadrature, is shown as the solid line.  Note that the
$i=3$ mode is ignored because of the smallness of its effect.}
\end{figure}
\clearpage


 
\section{Can matter clumping affect
the `mean' geometry of space?}

In spite of the feeble acoustic distortion induced by the time
delay perturbation of the primordial matter distribution, its
effect of biasing the acoustic peaks towards higher values of $\ell$
(as compared with the values under the scenario of a homogeneous Universe)
remains in principle an interesting phenomenon.   To illustrate,
let us repeat the acoustic smearing computation in the previous section
using a kernel $K(\theta,\ell)$ as determined by assuming a larger
normalization of the matter power spectrum (than the WMAP value
of Eq. (3)), viz.
\begin{equation} \delta\Phi \approx 0.057.  \end{equation}
The consequence, shown in Figure 6, is no longer negligible as before.
The systematic shift of the acoustic peaks towards higher $\ell$ (or smaller
$\theta$) is now evident, and tends to force the best-fit model parameters
to assume new values - that involve $\Omega <$ 1 in particular.
Thus clumping  does not just introduce local fluctuations in the metric
tensor (as first noted in section 3 of Einstein 1917), they
{\it can} mimic
space curvature, by
altering the mean geometry of space
as revealed by the statistical behavior of remotely emitted light
along many and varied
directions.  The magnitude of this influence is by no means necessarily
small: even the value of $\delta\Phi$ in Eq. (35) lies well within 
the perturbation criterion of $\delta\Phi \ll$ 1, yet it led to
a finite CMB $\delta\theta/\theta$ which, for the $i=1$ term of Eq. (1), is
{\it not} $\ll$ 1 because it
maxmizes at 32.5 \% when Eq. (31) is rescaled on the right side according
to $\delta\theta/\theta \sim (\delta\Phi)^2$.

In the light of the above development, let us
return to the question of why global space 
as measured by WMAP has zero curvature.
It has widely been accepted that
inflation solved the flatness problem, but does not
answer why $\delta\Phi$ is as small as the value in Eq. (3).  
which implies an unusually weak scalar-field coupling.  
This view is still correct, so long as
the `flatness' is understood as an intrinsic rather than
observed property of space, because indeed $\delta\Phi$, no
matter what values it assumes, does not alter the mean
intrinsic curvature.
But since,
as demonstrated in this work, even within the
perturbation regime a larger $\delta\Phi$ can make an
intrinsically flat geometry appear to WMAP as, on average, curved,
then one must acknowledge that inflation has not
explained why the {\it observed} 
mean curvature is zero.

\clearpage
\begin{figure}[H]
\begin{center}
\includegraphics[angle=90,width=4in]{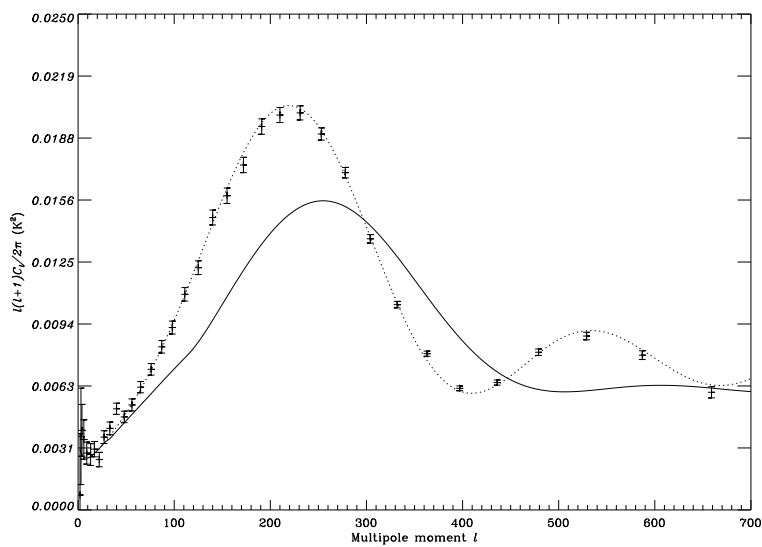}
\vspace{0cm}
\end{center}
\caption{{\it Under the assumption} of an enhanced spectral
normalization $\delta\Phi$
for the primordial density perturbation, i.e. with Eq. (35)
replacing Eq. (3), the {\it rescaled} (hence {\it mock}) WMAP3 data and
rescaled WMAP3 standard model prediction of the acoustic peaks
are plotted, with the latter as a dotted line.  This model is
then re-processed by the lensed correlation function due to time delay,
viz. Eq. (34) of section 5, 
again with the new $\delta\Phi$ value of Eq. (35).}
\end{figure}
\clearpage

\acknowledgements
The authors are grateful to Shaun Cole, Tom Kibble,
Chris Kochanek, and an anonymous 
referee for their most helpful advice.  RL thanks
Esra Bulbul for the preparation of Figures 2, 4, and 7.

\appendix
\section{APPENDIX}

\noindent
{\bf The basic relationship between gravitational potential perturbation
at various length scales and the matter power spectrum}

During some epoch $t$, or redshift $z$, let the typical random excursion
of the gravitational potential (from its mean value in a given cosmology)
be $\Phi_k$ at wavenumber $k$ or lengthscale
$2\pi a/k$.  By the Poisson equation, $\Phi_k$ is
expressible in terms of 
the matter density variation $\delta_k (z) = \delta\rho_m^k (z) /\rho_m (z)$ as
$\Phi_k = 4\pi G\rho_m (z) a^2 \delta_k (z) /k^2$, 
where $a=a(t) = 1/(1+z)$ is the expansion factor.
Squaring both sides, we obtain the
variance $\Phi_k^2$, the additive nature of which enables us 
to write the equation in
differential form, as
\begin{equation}  
d\Phi_k^2 = [4\pi G \rho_m (z) ]^2 a^4 \frac{d\delta_k^2}{k^4}.
\end{equation}
Under a matter dominated scenario, $\delta_k \propto (1+z)^{-1} \propto a$, 
and since
$\rho_m (z) = (1+z)^3 \rho_m (0) = \rho_m (0)/a^3$, we see that 
$d\Phi_k^2$ is independent of epoch.   Thus it
may be written in terms of
the present day quantities
\begin{equation} 
d\Phi_k^2 = [4\pi G \rho_m (0)]^2 \frac{d\delta_k^2 (0)}{k^4}, 
\end{equation}
where $d\delta_k^2 (0)$ relates to the $z=0$ matter power spectrum $P(k)$
as inferred from WMAP's standard model parameters (or, better still, the
WMAP1/2dFGRS external correlation, see below) by
\begin{equation} 
d\delta_k^2 (0) = \frac{dk}{k} \frac{k^3 P(k)}{2\pi^2}. 
\end{equation}
It should be noted, however, that if
vaccum energy (or dark energy) rather than matter dominates the low $z$
Universe, Eq. (A-2) would have slightly overestimated the value of
$\Phi_k$ at an earlier epoch.

Before proceeding further, therefore, 
we must estimate
a strict limit on the inaccuracy of Eq. (A-2).  Under a $\Lambda$-dominated
scenario, the rapid expansion freezes the density
distribution - the masses cannot move fast enough to counter it.  
As a result,
$\delta_k$ tends to a constant independent of $a$ (i.e. no further growth of
density contrasts in the linear regime), and
$\Phi_k$ decreases as $a^{-1}$.  In a $\Omega_m =$ 0.3, $\Omega_{\Lambda} =$
0.7, and $h =$ 0.7 cosmology (Bennett et al 2003, Spergel et al 2007), vacuum 
domination occurs at about $z \approx$ 0.32
(when $\Omega_m (1+z)^3 = \Omega_{\Lambda}$).  Thus, between $z \approx$ 0.32
and $z=$ 0, $d\Phi_k$ decreased from its hitherto constant value
by the fraction $1-a \approx$ 30 \%
(where $a$ refers to the expansion
factor at $z =$ 0.32).  It is clear then that any
lightpath integrations of the potential from $z =$ 0 back to some remote
past (or vice versa) will certainly not lead to an overestimation of $\Phi_k$
by more than 20 \%
if ones uses Eq. (A-2) for $\Phi_k$.

Let us {\it for the time being} suppose that the $z=0$
matter power spectrum has the simple form
\begin{equation} P(k) = Ak e^{-bk}, \end{equation}
so that at small $k$ the power spectrum takes the
Harrison-Zel'dovich form $P(k) \sim k$.
Then,
assembling Eqs. (A-2) and (A-3),
we have
\begin{equation} d\Phi_k^2 = 8 G^2 \rho_m^2 A e^{-bk} \frac{dk}{k} = 
(\delta\Phi)^2 A e^{-bk} \frac{dk}{k}, \end{equation}
where 
\begin{equation}
(\delta\Phi)^2 = \lim_{k\to 0} \frac{d\Phi_k^2}{d{\rm ln}k} =
\frac{9\Omega_m^2 H_0^4}{8\pi^2} A,
\end{equation}
with $H_0$ being the Hubble constant and $c =$ 1 here-and-after.
At sufficiently large $k$, corresponding to wavelengths smaller than
the size of the Universe (in today's distance scale)
during matter-radiation equipartition, $P(k)$ cuts off
because the modes could not grow.  Thus, in the simple manner by
which Eq. (A-4) depicted
the matter spectrum, $b \sim$ 33 Mpc, the equipartition horizon
for a flat Universe with $h =$ 0.7.  The advantage of Eq. (A-4), as 
can be seen in sections 3 and 4,
is that it affords us an analytical treatment of CMB foreground
re-processing by primordial matter - time delay and lensing in
particular - by a method of successive approximation from low to high
orders which reveals unamiguously the coherent spatial scales of
these effects.

\section{APPENDIX}

\noindent
{\bf Angular correlation function of
deflection due to primordial foreground matter}

The correlation function $C(\bth)$ as defined by the first equality of
Eq. (25), section 4, may be computed using Eqs. (24) and (11).  We find
 \begin{equation}
 \begin{array}{rl}\displaystyle
 C(\bth)=\fr{8(\de\ph)^2}{x^2}\int_0^xd\tilde x
 \int_{\frac{\tilde x}{2}}^{x-\frac{\tilde x}{2}}
 &d\bar x\fr{(x-\bar x)^2-\fr{1}{4}\tilde x^2}{r^4}\\
 & \times \left[2r^2-2\bar x^2\tha^2
 -\fr{b^2\bar x^2\tha^2}{b^2+r^2}
 +(3\bar x^2\tha^2-2r^2)\fr{b}{r}\arctan\left(\fr{r}{b}\right) \right],
 \end{array}
 \end{equation}
where $\bar x = (x'+x'')/2$, $\tilde x=x'-x''$ as in section 3, and
 \begin{equation} r^2=\tilde x^2+\bar x^2\tha^2. \end{equation}
because with the new $x$-variables we have, from
Eq. (12), $\br=(\tilde x,\bar x\bth)$.
By setting $\tha=0$ one arrives at the variance of the deflection angle
for a single light ray,
 \begin{eqnarray} C_0 = C(0) = \fr{\<\de\by^2\>}{x^2} &=&\fr{8(\de\ph)^2}{x^2}
\int_0^xd\tilde x
\int_{\frac{\tilde x}{2}}^{x-\frac{\tilde x}{2}}
 d\bar x\fr{(x-\bar x)^2-\fr{1}{4}\tilde x^2}{\tilde x^4}
 \left[2\tilde x^2-2b\tilde x
 \arctan\left(\fr{\tilde x}{b}\right)\right]\nonumber\\
 &=&\fr{8(\de\ph)^2}{3x^2}\int_0^x d\tilde x
 \fr{2x^3-3x^2\tilde x+\tilde x^3}{\tilde x^2}
 \left[1-\fr{b}{\tilde x}\arctan\left(\fr{\tilde x}{b}\right)\right]
 \end{eqnarray}
Evaluating the integral yields
 \begin{equation}
 C_0=C(0)=\fr{\<\de\by^2\>}{x^2}\ap\fr{4\pi}{3}(\de\ph)^2\fr{x}{b},
 \end{equation}
in the limit $x \gg$ b.

In general, the cross-correlation function $C(\tha)$ of Eq. (B-1) may
be expanded as a Taylor series:
 \begin{equation} C(\bth)=C_0+\half C_2\tha^2+\mathcal{O}(\tha^4), \end{equation}
where
 \begin{eqnarray}
 C_2 &=& \left[\frac{d^2}{d\tha^2} C(\tha) \right]_{\tha=0}\nonumber\\
 &=&\fr{32\de\ph^2}{x^2}\int_0^xd\tilde x
 \int_{\frac{\tilde x}{2}}^{x-\frac{\tilde x}{2}}
 d\bar x\fr{[(x-\bar x)^2-\fr{1}{4}\tilde x^2]\bar x^2}{\tilde x^4}
 \left[3\fr{b}{\tilde x}\arctan\fr{b}{\tilde x}
 -2-\fr{b^2}{\tilde x^2+b^2}\right]\nonumber\\
 &=&\fr{4\de\ph^2}{15x^2}\int_0^xd\tilde x
 \fr{4x^5-10x^3\tilde x^2+5x^2\tilde x^3+\tilde x^5}{\tilde x^4}
 \left[3\fr{b}{\tilde x}\arctan\fr{b}{\tilde x}
 -2-\fr{b^2}{\tilde x^2+b^2}\right].
 \end{eqnarray}
Again as before, in the limit $x \gg$ b only the leading term
needs to be kept.  The result is
 \begin{equation}
 C_2 = -\fr{2\pi}{15}(\de\ph)^2\left(\fr{x}{b}\right)^3.
 \end{equation}
We finally arrive at the expansion
 \begin{equation} 
 C(\bth)= \frac{\<\de y_i(\half\bth)\de y_i(-\half\bth)\>}{x^2} =
 \frac{4\pi}{3} (\delta\Phi)^2 \frac{x}{b} \left(1-\frac{1}{20}
\frac{x^2\tha^2}{b^2} + \cdots \right).
 \end{equation}
after substituting Eqs. (B-4) and (B-6) into Eq. (B-5).



\end{document}